\input harvmac
\input epsf
\newcount\figno
\figno=0
\def\fig#1#2#3{
\par\begingroup\parindent=0pt\leftskip=1cm\rightskip=1cm\parindent=0pt
\baselineskip=11pt
\global\advance\figno by 1
\midinsert
\epsfxsize=#3
\centerline{\epsfbox{#2}}
\vskip 12pt
{\bf Fig.\ \the\figno: } #1\par
\endinsert\endgroup\par
}
\def\figlabel#1{\xdef#1{\the\figno}}
\def\encadremath#1{\vbox{\hrule\hbox{\vrule\kern8pt\vbox{\kern8pt
\hbox{$\displaystyle #1$}\kern8pt}
\kern8pt\vrule}\hrule}}

\lref\rY{``{\it Holographic Reformulation of string Theory in the pp Wave 
Limit}", S. Dobashi, H. Shimada 
and T. Yoneya, hep-th/0209251.}
\lref\rBMN{``{\it Strings in Flat Space and pp Waves from N=4 Super 
Yang-Mills}", 
D. Berenstein, J. Maldacena, H. Nastase, {\it JHEP} {\bf 0204}:013, (2002), 
hep-th/0202021.} 

\lref\rKPSS{``{\it A New Double-Scaling Limit of N=4 Super Yang-Mills and 
PP-Wave Strings},"
C. Kristjansen, J. Plefka, G.W. Semenoff and M. Staudacher, hep-th/0205033\semi
``{\it BMN correlators and Operator Mixing 
in N=4 super Yang-Mills Theory},"
N.Beisert, C. Kristjansen, J. Plefka, G.W. Semenoff and M. Staudacher, 
hep-th/0208178.}

\lref\rLMP{``{\it Cubic Interactions in PP-Wave Light Cone String Field 
Theory,}"
P. Lee, S. Moriyama and J. Park, hep-th/0206065\semi
``{\it A Note on Cubic Interactions in PP-Wave Light Cone String Field Theory,"} 
P.Lee, S. Moriyama and J. Park, hep-th/0209011.}

\lref\rM{``{\it The Large $N$ Limit of Superconformal Field Theories and 
Supergravity},"
J.M. Maldacena, Adv. Theor. Math. Phys. {\bf 2} 231, (1998), hep-th/9711200.}

\lref\rGKP{``{\it Gauge Theory Correlators from Noncritical String Theory},"
S.S. Gubser, I.R. Klebanov and A.M. Ployakov, Phys. Lett. {\bf B428} 105 
(1998).}

\lref\rGKPtwo{``{\it A Semiclassical Limit of the Gauge/String Correspondence},"
S.S. Gubser, I.R. Klebanov and A.M. Ployakov, hep-th/0204051.}

\lref\rW{``{\it Anti-de Sitter Space and Holography,}" E. Witten, Adv. Theor. 
Math. Phys.
{\bf 2} 253 (1998), hep-th/9802150.}

\lref\rGMR{``{\it Operators with Large R Charge in N=4 Yang-Mills Theory}," D.J. 
Gross,
A. Mikhailov and R. Roiban, hep-th/0205066.}

\lref\rSZ{``{\it Exact Anomalous Dimensions of N=4 Yang-Mills Operators with 
large R charge},"
Alberto Santambrogio and Daniela Zanon, hep-th/0206079.}

\lref\rCKT{``{\it Three-point functions in N=4 Yang-Mills Theory and pp-waves},"
C.S. Chu, V.V. Khoze and G. Travglini, hep-th/0206005.}

\lref\rT{``{\it Reformulating  String Theory With The
    1/N Expansion},"
    Sakharov Conf.1991:0447-454 (QC20:I475:1991)
    In *Moscow 1991, Proceedings, Sakharov memorial lectures in
    physics, vol. 1* 447-453, C.Thorn, hep-th/0205066.}

\lref\rHT{``{\it Large N Matrix Mechanics on  the Light Cone}," 
M. Halpern and C. Thorn, hep-th/0111280.}

\lref\rBT{``{\it World Sheet Description of Large N Quantum Field Theory}," 
K. Bardakci and C. Thorn, hep-th/0110301.}

\lref\rThooft{``{\it A Planar Diagram Theory for Strong Interactions}," G. 't 
Hooft, 
{\it Nucl. Phys.} {\bf B72} 461, (1974).} 

\lref\rPolya{``{\it String Representations and Hidden Symmetries for Gauge 
Fields}," A. M. 
Polyakov, {\it Phys. Lett.} {\bf 82B} 247 (1979).}

\lref\rSakita{``{\it Field Theory of Srings as a Collective Field Theory of U(N) 
Gauge
Field}," B. Sakita, {\it Phys. Rev.} {\bf D21} 1067 (1980).}

\lref\rBN{``{\it On Light Cone String Field Theory from Superyang-Mills and 
Holography},"
D. Berenstein and H. Nastase, hep-th/0205048.}

\lref\rMIT{``{\it pp Wave String Interactions from Perturbative Yang-Mills 
Theory}," N. R. 
Constable, D Freedman, M. Headrick, S. Minwalla, L. Motl, A. Postnikov,  W. 
Skiba {\it JHEP}
{\bf 0207}:017, (2002), hep-th/0205089.}

\lref\rSV{``{\it Superstring Interactions in a pp-wave Background}," 
M. Spradlin and A. Volovich, hep-th/0204146\semi
``{\it Superstring Interactions in a pp-wave Background II}," 
M. Spradlin and A. Volovich, hep-th/0206073.}

\lref\rSbook{``{\it Quantum Theory of Many Variable Systems and Fields}," B. 
Sakita,
World Sci, Let. Notes Phys.1:1-217 (1985).}

\lref\rJS{``{\it The Quantum Collective Field method and Its Application to the 
Planar Limit},"  
A. Jevicki, B. Sakita, {\it Nucl. Phys.} {\bf B165} 511, (1980).} 

\lref\rOP{``{\it Loop Space Representation and the Large N behavior of the one 
plaquette
Kogut-Susskind Hamiltonian}," A. Jevicki and B. Sakita, {\it Phys. Rev.} {\bf 
D22} 467 (1980).}

\lref\rDJ{``{\it String Field Theory and Physical Interpretation of D=1 
Strings}," 
S. R. Das, A. Jevicki, {\it Mod. Phys. Lett.} {\bf A5} 1639 (1990).} 

\lref\rJR{``{\it Loop Space Hamiltonians and Field Theory of Noncritical 
Strings},"
A. Jevicki, J. P. Rodrigues, {\it Nucl. Phys.} {\bf B421} 278 (1994).} 

\lref\rDJR{``{\it Scattering Ampltiudes and Loop Corrections in Collective 
String Field Theory},"
K. Demeterfi, A. Jevicki, J. P. Rodrigues, {\it Nucl. Phys.} {\bf B362} 173 
(1991); 
{\bf B365} 499, (1991).}

\lref\rMaster{``{\it Glueballs, Mesons, Surface Roughening and the Collective 
Field Method},"
I. Affleck, {\it Proceedings of the Brown University Workshop on 
Non-perturbative Studies in QCD}, edited by A. Jevicki, C.-I Tan, Brown Report 
No. HET-457
(1981), (unpublished) \semi
``{\it Loop Space Master Variables and the Spectrum in the Large N Limit},"
J.P. Rodrigues, PhD thesis, Brown University (1983) \semi
``{\it Master Variables and Spectrum Equations in Large N Theories},"
A. Jevicki, J. P. Rodrigues {\bf B230 [FS10]} 317 (1984).}

\lref\rG{``{\it String Interactions in PP-waves}," R. Gopakumar, 
hep-th/0205174.}

\lref\rKKLP{``{\it pp-wave/Yang-Mills Correspondence: An explicit Check},"
Y.J. Kiem, Y.B. Kim, S.M. Lee and J.M. Park, hep-th/0205279.}

\lref\rH{``{Three point functions of N=4 Super Yang-Mills from light cone string 
field theory
in pp-wave}," M.X. Huang, hep-th/0205311.}

\lref\rV{``{\it Bits, Matrices and 1/N}," H. Verlinde, hep-th/0206059.}

\lref\rKSV{``{\it New Effects in Gauge Theory from pp-wave Supertsrings},"
I.R. Klebanov, M. Spradlin and A. Volovich, hep-th/0206221.}

\lref\rJStwo{``{\it Collective Field Approach to the Large-N Limit: 
Euclidean field theories},"
A. Jevicki, B. Sakita, {\it Nucl. Phys.} {\bf B185} 89 (1981).}

\lref\rSY{``{\it Stochastic Hamiltonians for Noncritical String Field Theories 
from Double Scaled Matrix Models}," F.Sugino and T.Yoneya, {\it Phys.Rev.} {\bf 
D53} 
4448-4488 (1996), hep-th/9510137.}

\lref\rIK{``{\it String Field Theory of $c\le 1$ Noncritical Strings}," 
N.Ishibashi and H.Kawai, 
{\it Phys.Lett.} {bf B322} 67 (1994).}

\lref\rJRAM{``{\it Noncommutative Gravity from the AdS/CFT Correspondence}," 
Antal Jevicki and Sanjaye Ramgoolam, {\it JHEP} {\bf 9904}:032, (1999),
hep-th/9902059.}

\lref\rRAJEEV{``{\it Collective Potential for Large N Hamiltonian Matrix Models 
and Free
Fischer Information}," A. Agarwal, L. Akant, G.S. Krishnaswami and S.G. Rajeev,
hep-th/0207200.}

\lref\rS{``{\it Comments on Superstring Interactions in a Plane-Wave Background}," 
J.H.Schwarz, hep-th/0208179.}

\lref\rFM{``{\it Operator Mixing and the BMN Correspondence}," N. Constable, 
D. Z. Freedman, M. Headrick 
and S. Minwalla, hep-th/0209002.}
\lref\rDDJR{``{\it Derivation of String Field Theory from the Large N BMN Limit},"
R.de Mello Koch, A.Donos, A.Jevicki
and J.P.Rodrigues, hep-th/0305042.}


\lref\rMM{Yu.M. Makeenko, A.A. Migdal, 
``{\it Exact Equation for the Loop Average in Multicolor QCD}," 
Phys. Lett. {\bf B88} (1979) 135.}


\Title{ \vbox {\baselineskip 12pt\hbox{}
\hbox{}  \hbox{September 2002}}}
{\vbox {\centerline{Collective String Field Theory of Matrix Models} 
\centerline{in the BMN Limit}
}}

\smallskip
\centerline{Robert de Mello Koch$^\dagger$, Antal Jevicki$^*$ and Jo\~ao P. 
Rodrigues$^\dagger$}
\smallskip
\centerline{\it Department of Physics and Center for Theoretical 
Physics$^\dagger$,}
\centerline{\it University of the Witwatersrand,}
\centerline{\it Wits, 2050, South Africa}
\centerline{\tt robert,joao@neo.phys.wits.ac.za}
\smallskip
\centerline{\it Department of Physics$^*$,}
\centerline{\it Brown University,}
\centerline{\it Providence, RI 02912, USA}
\centerline{\tt antal@het.brown.edu}
\bigskip

\centerline{\it This work is dedicated to the memory of Bunji Sakita.}

{\vskip 20pt}

\noindent
We develop a systematic procedure for deriving canonical string field theory 
from large N matrix models in the Berenstein-Maldacena-Nastase limit. The 
approach, based on collective field theory, provides a generalization of
standard string field theory.


\Date{}

\def\MAKEdalamSIGN#1#2{%
\setbox0=\hbox{$\mathsurround 0pt #1{#2}$}
\dimen0=\ht0 \advance\dimen0 -0.8pt
\hbox{\vrule\vbox to\ht0{\hrule width\dimen0 \vfil\hrule}\vrule}}


\newsec{Introduction}
Berenstein, Maldacena and Nastase (BMN) have established a correspondence 
between
the spectrum of closed superstrings in the pp-wave background and ${\cal N}=4$ 
super Yang-Mills 
theory \rBMN. In the process, they defined a (double scaling) limit of 
Yang-Mills theory 
with large N and large R-charge in which string theory is to be obtained. This 
limit 
simultaneously simplifies and extends the previous AdS/CFT correspondence in 
which the map 
was understood for supergravity states \rM,\rGKP,\rW . It represents a 
significant step 
in establishing the long held expectation that large $N$ gauge theories at 
nonperturbative 
level lead to string theory \rThooft,\rPolya,\rSakita,\rMM,\rJStwo,\rT,\rBT,\rHT. The 
new 
correspondence has the 
promise of providing a gauge theoretic description of string theory in a variety 
of backgrounds. 

Vigorous studies are now being carried out in establishing \rGKPtwo,\rKPSS,\rGMR 
,\rSZ,\rCKT
the Berenstein, Maldacena and Nastase correspondence. Much of the focus 
represents an effort 
to understand the interactions
generated by Yang-Mills theory. Positive results but with certain limitations 
have been 
obtained in the literature \rSV,\rBN,\rMIT,\rG ,\rKKLP,\rH,\rLMP, \rV,\rKSV,\rS. 
One of the 
difficulties is that short of direct comparison of amplitudes (between pp-wave 
strings and YM correlators) no systematic procedure was developed. 
The comparison in question has been 
limited at the present time to a set of correlators termed perturbative, but for 
a much 
larger set of so-called non-perturbative correlators, it still remains to be 
established.
Related are issues of unitarity \rFM\  for the light-cone type field theory.
A proposal for a holographic map that was given in \rY.  

In this work we formulate a systematic scheme of obtaining canonical string 
field theory in the 
BMN limit. The method we apply is that of collective field theory 
\rSakita,\rJS,\rSbook , which provides a 
clear and well defined scheme for making the transition from YM to a string 
theory 
description. The method was employed successfully in establishing the first 
matrix model/string 
field theory map, namely that of non-critical strings \rDJ, \rDJR.
By its nature the method represents a direct change of variables from the
matrices of U(N) gauge theory to the fields of string theory. We focus in this 
paper on generic matrix models  for 
notational reason as 
they are sufficient in demonstrating the basics of the present method. The 
simplest example of the derivation
of string type interactions is given already in the free model. For 
SUGRA type amplitudes this
essentialy gives the full result. The effect of YM interactions and $g_{YM}$ 
corrections
will be given the next paper\rDDJR.

Our plan is as follows:
After the Introduction in Sect.1, we summarize the basics of collective
field theory in Sect.2. In Sect.3 we give a simple example with permutation 
symmetry. In this case
we discuss the passage from time-like gauge field theory to the light cone which 
is of relevance 
when performing a comparison with previous works. In Sect.4 we discuss the free 
matrix model in 
the BMN limit, concentrating on supergravity type modes. For these, Yang-Mills 
type interactions
do not play a significant role and we exhibit agreement with SFT already at this 
level. In sect.5 we
discuss general
'stringy states' exhibiting the corresponding 3-string interaction.

In Sect.6 under the conclusions,
we discuss the generality of collective string field theory (CSFT) in comparison 
with
standard light cone field theory.

\newsec{Collective field theory}

In this section, we give a general overview of collective field theory
\rJS,\rSbook. It represents a systematic formalism for describing the dynamics 
of invariant
observables of the theory. In gauge or matrix theory the physical observables 
are given by
loops or traces of matrix products (words). The method consists of a direct 
change of variables
to the invariant observables. The result is a (collective) Hamiltonian 
describing the full 
dynamics of the theory.

To be specific, let us consider a complex multi-matrix system with Hamiltonian 

\eqn\Example
{H = - \Tr(\sum_{i=1}^M {\partial\over\partial \bar{Z}_i}{\partial\over\partial 
{Z}_i}) 
+ V (\bar{Z}_i,Z_i).} 

\noindent 
where the potential $V (\bar{Z}_i,Z_i)$ is invariant under 

$$ Z_i \to U^{\dagger} Z_i U \qquad   
\bar{Z}_i \to U^{\dagger} \bar{Z}_i U .$$

\noindent
Then, the Hamiltonian is invariant under the above symmetry, and one
may consider equal time single trace correlators (operators) of the 
form

\eqn\Word{\Tr (. . .  
\prod_{i=1}^M Z_i^{n_i} {\bar{Z}_i}^{\bar{n}_i}    
\prod_{j=1}^M Z_j^{m_j} {\bar{Z}_j}^{\bar{m}_j} 
. . .).}

In the large $N$ limit, the change of variables from the original 
variables to loop variables implies a reduction of degrees of 
freedom. For instance, in atomic systems or single matrix systems
the collective (loop) variables correspond to the
radial component and eigenvalue basis, respectively.
There is by now ample evidence coming both from studies of single 
matrix models \rOP\ and matrix descriptions of lower dimensional strings \rDJ, 
\rDJR,\rIK,\rJR,\rSY\
that these variables can be treated as independent in the large N limit. 
Possible constraints 
due to the finite size of matrices can always be imposed after the change of 
variables. This 
results in interesting effects related to the large N exclusion principle 
\rJRAM. 

Let us denote the invariant field variables by 

$$ \phi_{C} $$

\noindent
where ${C}$ is a loop or word index. For simplicity of notation, 
we require this index to include $ \bar{\phi}_{C} $. One may then 
consider changing variables from the original variables to this new set 
\rJS. Concentrating on the kinetic term, one has:

\eqn\Kinetic{
T = - \Tr(\sum_{i=1}^M {\partial\over\partial \bar{Z}_i}
{\partial\over\partial {Z}_i}) = 
-\sum_{C,C'} \Omega (C,C') {\partial\over\partial \bar{\phi}_C}
{\partial\over\partial {\phi}_{C'}} +
\sum_{C}  \omega (C) {\partial\over\partial {\phi}_{C}} 
}

\noindent  
where

$$  \Omega (C,C') = \Tr(\sum_{i=1}^M 
{\partial\bar{\phi}_C\over\partial\bar{Z}_i}{\partial{\phi}_{C'}\over\partial 
{Z}_i})
= \bar{\Omega}(C',C) 
$$

\noindent
and 

$$
\omega (C) = - \Tr(\sum_{i=1}^M {\partial^2\phi_{C}\over\partial 
\bar{Z}_i\partial {Z}_i})
$$

\noindent
$\Omega (C,C')$ ``joins" loops, or words. As an example, if 
$\phi_{C} = \Tr (Z_1^J)$ and $\phi_{C'} = \Tr (Z_1^{J'})$ then
$\Omega = J {J'} \Tr (Z_1^{J-1} \bar Z_1^{J'-1})$. So in general, one may write 
schematically 

$$   \Omega (C,C') =  \sum \phi_{C+C'}   $$

\noindent
where ${C+C'}$ is obtained by adding the two words $C$ and $C'$.
Similarly, $\omega$ ``splits" loops. Schematically again, 

\eqn\littleo{\omega (C) = \sum  \phi_{C'}\phi_{C''} }

\noindent
represents all the processes of splitting the word $C$ into 
$C'$ and $C''$.

As any change of variables the present one also involves a nontrivial Jacobian 
contributing to 
the measure. The Jacobian plays a central role in guaranteeing the unitarity of 
the new 
description. A most relevant aspect in the construction of collective field 
theory is the 
determination of the Jacobian. By performing a similarity transformation 

$$        {\partial\over\partial {\phi}_{C}} \to 
             J^{1/2} {\partial\over\partial {\phi}_{C}}  J^{-{1/2}} =
 {\partial\over\partial {\phi}_{C}} - {1\over 2}{\over}
{\partial\ln J\over\partial {\phi}_{C}} 
$$

\noindent
one imposes the requirement that the Hamiltonian \Kinetic\ becomes explicitly 
hermitean. This 
requirement of hermiticity provides a system of equations for the Jacobian. This 
set of 
equations establishes the necessary data for a complete rewriting of the 
hamiltonian in terms 
of the new observables. The differential equation is

\eqn\Jacobian{
 - (\bar{\partial}_{C'} \ln J) \tilde\Omega (C',C) = \omega (C) 
+ \bar{\partial}_{C'} \tilde\Omega (C',C)
}

\noindent 
with 

$$
\tilde\Omega(C,C') = {1 \over 2} \Big[ \Tr(\sum_{i=1}^M
{\partial\bar{\phi}_C\over\partial\bar{Z}_i}{\partial{\phi}_{C'}\over\partial 
{Z}_i})
+ \Tr(\sum_{i=1}^M
{\partial{\phi}_{C'}\over\partial\bar{Z}_i}{\partial\bar{\phi}_{C}\over\partial 
{Z}_i})\Big]
= \tilde\Omega(\bar{C'},\bar{C})
$$

\noindent
The explictly hermitean hamiltonian (collective field hamiltonian) is then
seen to read

$$
H =  ( {\partial\over \partial\bar{\phi}(C)}  + 
{1\over 2} {\partial\ln J\over\partial \bar{\phi}_{C} } ) \tilde\Omega(C,C') 
( - {\partial\over \partial\phi(C')} + 
{1\over 2} {\partial\ln J\over\partial {\phi}_{C'} } )     +  V        
$$

\noindent
Once use is made of \Jacobian, the leading contribution to the collective field 
hamiltonian
is then

\eqn\CFH{
H'  =  \Big(   - {\partial\over\partial \bar{\phi}_{C}} \tilde\Omega(C,C')
{\partial\over\partial {\phi}_{C'}}   + 
{1\over 4} \omega(C) \tilde\Omega^{-1}(C,C')\bar\omega(C') \Big) + V}

\noindent
or 

$$
H'= \Big(    \bar\Pi(C) \tilde\Omega(C,C') \Pi(C') +
{1\over 4} \omega(C) \tilde\Omega^{-1}(C,C')\bar\omega(C')\Big) + V
$$

\noindent
where 
$$
\Pi(C) = - i {\partial\over \partial\phi(C)} \qquad 
\bar\Pi(C) = - i {\partial\over \partial\bar{\phi}(C)}
$$

The full Hamiltonian in addition contains counterterms which contribute at loop 
level.  Their form is explicitly determined by the hermiticity requirement and 
reads

$$
{\eqalign{\Delta H &= - {1\over 2} \, {\partial \omega(C)\over\partial \phi_C}
+ {1\over 2} \, {\partial \tilde{\Omega}(C^{''},C')\over \partial \phi_{C^{''}}} 
\, \tilde{\Omega}^{-1}(C',C) 
\bar\omega (C) \cr
&+{1\over 4} \, {\partial\tilde{\Omega}(C^{''},C)\over \partial \phi_{C^{''}}} 
\, \tilde{\Omega}^{-1}(C,C') \, 
{\partial \tilde{\Omega} (C',C^{'''})\over \partial\phi_{C^{'''}}}}}$$

Let us now comment on some relevant features of the collective Hamiltonian. 
First of all any 
matrix model interaction will appear in this approach as a tadpole term (ie 
linear in the
fields).
As such they will have a role in the full theory (specific backgrounds 
correspond to minima of the effective potential 
${1/4} \,\omega \tilde{\Omega}^{-1} \bar\omega + V$ ) but the basic non 
linearity is seen
to emerge from the free, kinetic term. These will be 'renormalized' by the 
matrix interactions.
One also sees that in addition to simple cubic interactions,
the formalism generates a sequence of higher point interactions(vertices). These 
come from the expansion
of ${1/4}\,\omega \tilde{\Omega}^{-1} \bar\omega $, the non trival effective 
potential of collective 
field theory (for a
recent study of this term see \rRAJEEV). In addition one has the loop 
counterterms.
All these are specified by the requirement of Hermiticity. The Hamiltonian \CFH\ 
is sufficient to describe fluctuations
about a given background \rMaster. Furthermore, as it will be described in the 
following, the term 
$\Pi\Omega\Pi$ will contain enough information to study the structure of three 
point interactions.

The non triviality in applying collective field theory to multimatrix models or 
Yang-Mills theory
is contained in the following. In the context of interest to us
in this paper, BMN have identified a set 
of observables (traces)
which have a mapping into the pp wave string. If in the collective approach one 
begins with 
a  given
set of loops  $C$ and their conjugates $C'$, through the process of ``joining" 
(contained in $\Omega (C,C')$) one generates 
new loops not in the original set. It is this sequence of extra 
degrees of freedom that we have to understand
in the present approach. Their relevance will also point to a more 
general scheme contained in collective string field theory(CSFT).
 
\newsec{ Example With $S_N$ Symmetry}

In this section a simple system of particles with $S_N$ symmetry will be given. 
This example is 
to illustrate the strategy we employ in our study of matrix models, for which 
the corresponding 
collective field theories are not yet explicitly available. Apart from 
illustrating the 
workings of collective field theory in this example we establish some facts 
about the large $J$ 
limit which will be of general value. Start with the free
N-body Hamiltonian 



\eqn\ToyH
{H=-2\sum_{i=1}^N{\partial\over\partial z_i}
{\partial\over\partial\bar{z}_i}+{\omega^2\over 2}
\sum_{i=1}^N\bar{z}_i z_i .}

\noindent
where

$$ z_i=x_i+iy_i,\qquad \bar{z}_i=x_i-iy_i ,$$

\noindent
The relevant collective field variables are the $S_N $ invariants

\eqn\CollF
{\phi_{nm} =\sum_{i=1}^N\, \bar{z}_i^n \, z_i^m.}

\noindent  After a Fourier transform, they become the density
 fields $\, \phi(\bar{Z} , Z)$.  The change of variable results in the

\eqn\CollH
{\eqalign{H&=\int dx dy\Big(2\partial_{\bar{z}}\Pi\phi \partial_z\Pi +
{1\over 2}{\partial_{\bar{z}}\phi\partial_z\phi\over\phi}+
{1\over 2}\omega^2 z\bar{z}\phi \Big)\cr
&\equiv \int dx dy 2\partial_{\bar{z}}\Pi\phi \partial_z\Pi 
+{\cal V}\big[\phi\big].}}

\noindent
The expansion of this Hamiltonian about the leading large $N$ configuration
will generate an infinite number of interaction vertices, despite the fact
that the original system is noninteracting. The leading large
$N$ configuration

\eqn\LeadC
{\phi_0 ={N\omega\over\pi}e^{-\omega z\bar{z}},}

\noindent
minimizes the effective potential ${\cal V}\big[\phi\big]$. Expanding
about the leading configuration

\eqn\scale
{\Pi ={1\over \sqrt{\phi_0}}\tilde{\Pi},\qquad 
\phi =\phi_0+\sqrt{\phi_0}\eta,}

\noindent
we obtain the following quadratic Hamiltonian

\eqn\Quad
{H_2={1\over 2}\int dx dy\Big(
\tilde{\Pi}(-4\partial_z\partial_{\bar{z}}
+\omega^2 z\bar{z}-2\omega)\tilde{\Pi}
+\eta (-4\partial_z\partial_{\bar{z}}
+\omega^2 z\bar{z}-2\omega)\eta\Big).}

\noindent
We now expand $\tilde{\Pi}$ and $\eta$

\eqn\Expnd
{\eta= \sum_{J=1}^\infty \Big(\bar{c}_J\psi_J +c_J\bar{\psi}_J
\Big),\qquad
\tilde{\Pi}= \sum_{J=1}^\infty \Big(p_J
\psi_J +\bar{p}_J\bar{\psi}_J\Big)}

\noindent
in terms of the Harmonic oscillator eigenfunctions

\eqn\EigenModes
{\psi_J= \sqrt{\omega^{J+1}\over \pi J!}z^J 
e^{-{\omega\over 2}z\bar{z}},\qquad 
\bar{\psi}_J= \sqrt{\omega^{J+1}\over \pi J!}\bar{z}^J 
e^{-{\omega\over 2}z\bar{z}},}

\noindent
which satisfy

\eqn\EigenEqn
{\eqalign{(-4{\partial\over\partial z}{\partial\over\partial\bar{z}}
+\omega^2 z\bar{z})\psi_J &= 2(J+1)\omega\psi_J,\cr
(-4{\partial\over\partial z}{\partial\over\partial\bar{z}}
+\omega^2 z\bar{z})\bar{\psi}_J &= 2(J+1)\omega\bar{\psi}_J.}}

\noindent
The quadratic Hamiltonian becomes

\eqn\QuadH
{H_2={1\over 2}\int dx dy\Big(J\omega p_J\bar{p}_J+J\omega c_J\bar{c}_J\Big).}

\noindent
Notice that $H_2$ gives the exact spectrum of our toy model.

We now consider the interactions generated by the collective field
theory. There are two contributions to the cubic vertex.
First we have the momentum field dependent term

$$ \int dx dy 2\partial_z\Pi\sqrt{\phi_0}\eta \partial_{\bar{z}}\Pi $$

\noindent
giving

$$\sum_{J_1,J_2,J_3}4\pi{\cal M} \int dr\,\,\, r^{J_1+J_2+J_3-1}
e^{-\omega r^2}(\bar{c}_{J_2}\bar{p}_{J_3}p_{J_1}+
c_{J_2}\bar{p}_{J_1}p_{J_3})\delta_{J_2+J_1,J_3}.$$

$${\cal M}^2={N\omega^{J_1+J_2+J_3+2}\over\pi^2 N^2 J_1!J_2!J_3!}.$$

\noindent
After integration over $r$ we obtain

$$ V_3=\sum_{J_1,J_2,J_3} \omega
({J_1+J_2+J_3\over 2}-1)!
{2J_1 J_3\over \sqrt{NJ_1!J_2!J_3!}}
(\bar{p}_{J_3}\bar{c}_{J_2}p_{J_1}+\bar{p}_{J_1}p_{J_3}c_{J_2})
\delta_{J_1 +J_2,J_3} .$$

\noindent
In an identical way one evaluates the three point interaction coming from
the $\omega\Omega^{-1}\omega ={\partial_{\bar{z}}\phi\partial_z\phi\over\phi}$ 
term.
When expanded in terms of the fluctuating field this term generates an infinite 
series of vertices. At the cubic level it gives:

$$ \tilde{V}_3=-\sum_{J_1,J_2,J_3} {\omega\over 16}
({J_1+J_2+J_3\over 2}-1)!
{(J_1+J_2+J_3)^2\over \sqrt{NJ_1!J_2!J_3!}}
(\bar{c}_{J_1}\bar{c}_{J_2}c_{J_3}+\bar{c}_{J_3}c_{J_1}c_{J_2})
\delta_{J_1 +J_2,J_3} .$$

At this point we pause and discuss the result and make some relevant comments. 
The 
collective hamiltonian that we have constructed is a hamiltonian in a time-like 
gauge. 
The (collective) fields $c_J$ and $\bar {c} _J$ represent left and right 
movers (with J being a momentum). These  modes correspond to observables with a 
pure $z^J$ 
or a $\bar{z}^J$ dependence. In addition we have the fields  $p_J$ and 
$\bar{p}_J$ which correspond to
an insertion of a time derivative of $Z$. Studies of string theory in the 
pp-wave background 
discuss construction of a light cone hamiltonian (it was in fact notoriously 
difficult to 
construct interacting string field theory 
hamiltonians in a time like gauge). From our time-like hamiltonian 
we can pass to a light cone one by taking the infinite momentum (large J) limit. 
Introduce 
creation-annihilation operators as usual

\eqn\Osc
{\eqalign{c_J&=a_J+\tilde{a}_J^\dagger ,\qquad
\bar{c}_J=a_J^\dagger +\tilde{a}_J,\cr
p_J&=-{i\over 2}(a_J^\dagger -\tilde{a}_J),\qquad
\bar{p}_J={i\over 2}(a_J -\tilde{a}_J^\dagger )}}

\noindent
Passage to the lightcone frame is now implemented by dropping the $\tilde{a}$, 
$\tilde{a}^\dagger$
oscillators. This represents a reduction of degrees of freedom. Reducing the 
$\Pi\Omega\Pi =\partial_{\bar{z}}\Pi\phi\partial_z\Pi$ term 
contribution to just right movers gives

$$\sum_{J_1,J_2,J_3} {\omega\over 4}
\sqrt{J_3!\over NJ_1!J_2!}(J_1 + J_2)
(a_{J_1}^\dagger a_{J_2}^\dagger a_{J_3}+a_{J_3}^\dagger a_{J_1}a_{J_2})
\delta_{J_1 +J_2,J_3} .$$

\noindent
Our second interaction term gives

$$-\sum_{J_1,J_2,J_3} {\omega\over 4}
J_3\sqrt{J_3!\over NJ_1!J_2!}
(a_{J_1}^\dagger a_{J_2}^\dagger a_{J_3}
+a_{J_3}^\dagger a_{J_1}a_{J_2})
\delta_{J_1 +J_2,J_3} .$$

\noindent
Summing these two contributions we obtain the light cone cubic vertex 

\eqn\CubV
{H_3=\sum_{J_1,J_2,J_3} {1\over 4}
\sqrt{J_3!\over NJ_1!J_2!}(J_1\omega + J_2\omega - J_3\omega)
(a_{J_1}^\dagger a_{J_2}^\dagger a_{J_3}+a_{J_3}^\dagger a_{J_1}a_{J_2})
\delta_{J_1 +J_2,J_3} .}

\noindent
Here we notice that after summing  the resulting vertex comes with a prefactor
equal to the difference in energies of the participating modes. 
Thus this vertex is seen to vanish on shell. We also see that the prefactor
responsible for this is the one proposed in \rMIT. We emphasize that our 
full timelike gauge interaction does not vanish (on shell), it is only after the
projection to light cone frame that we obtain the vanishing prefactor.
This cancellation came from the two types of terms and 
represents a general property of  
collective field interaction when projected to the infinite momentum frame.

It is not obvious how this vertex could 
possibly reproduce the three point functions of the original model. To 
explore this point, note that equal time correlation functions of $z_i$ and 
$\bar{z}_i$ map into the equal time correlation functions of modes of the 
collective field

\eqn\CorrFncs
{\eqalign{\langle\big(&\int d^2 x_1 z_1^{J_1}\phi(x_1)\big)
\big(\int d^2 x_2 z_2^{J_2}\phi(x_2)\big)
\big(\int d^2 x_3 \bar{z}_3^{(J_1 +J_2)}\phi(x_3)\big)\rangle\cr 
&= \langle\sum_{i=1}^N (x_i +iy_i)^{J_1}
\sum_{j=1}^N (x_j +iy_j)^{J_2}
\sum_{k=1}^N (x_k -iy_k)^{J_1+J_2} \rangle .}}

\noindent
The correlators in the original model are easily evaluated

\eqn\CorrOne
{\langle\sum_{i=1}^N (x_i +iy_i)^{J_1}
\sum_{j=1}^N (x_j +iy_j)^{J_2}
\sum_{k=1}^N (x_k -iy_k)^{J_1+J_2} \rangle =
N{(J_1 +J_2)!\over \omega^{J_1 +J_2}}.}

\noindent
Now consider the collective field theory calculation. Performing
the integrations

$$\eqalign{\langle\big(&\int d^2 x_1 z_1^{J_1}\phi(x_1,t)\big)
\big(\int d^2 x_2 z_2^{J_2}\phi(x_2,t)\big)
\big(\int d^2 x_3 \bar{z}_3^{(J_1 +J_2)}\phi(x_3,t)\big)\rangle \cr
&=\sqrt{N^3 J_1! J_2! J_3! \over\omega^{J_1 +J_2 +J_3}} 
\langle A_{J_1} A_{J_2} A_{J_3}^\dagger (t)\rangle .}$$

\noindent
where we have the dressed creation-anihilation operators. In terms of
 first order perturbation
theory these receive a contribution given by the cubic interactions

$$ {1\over 4}\sqrt{J_3!\over N J_1! J_2!}
(\omega J_1 +\omega J_2 -\omega J_3) $$

\noindent
multiplied by the
(propagators) factor

$${2\over \omega J_1+ \omega J_2- \omega J_3 },$$

\noindent
which neatly cancels the energy prefactor contained in the cubic vertex
reproducing the correlator \CorrOne.

In this simple toy model, we see that the structure of the light cone cubic 
vertex is seen 
by the $\Pi\Omega\Pi$ term of the collective field theory Hamiltonian. The
net effect of the $\omega\Omega^{-1}\omega$ term is to complete the cubic vertex
so that it comes multiplied by the correct prefactor. Thus, to compute the 
vertex 
up to the prefactor we would only need to consider the $\Pi\Omega\Pi$ term. This
appears to be a rather general conclusion. The collective field theory also 
provides
a natural explanation of the difference in energies prefactor multiplying the 
cubic 
vertex. 

\newsec{Collective String Field Theory of Matrix Models}

The BMN study addressed the ${\cal N}=4$ Super Yang-Mills theory in 
four dimensions. In this particular case
arguments were presented for a correspondence with string theory 
in the ppwave background. Through investigation of
correlation functions of the theory it became visible that some of the basic 
structure is carried by the matrix (Higgs) 
fields of the theory. It is these degrees of freedom that we will 
concern ourselves with in our study. Clearly the full
(Yang-Mills) theory is relevant for the actual correspondence and our discussion 
of the present 
matrix model (done
for notational simplicity) should be viewed in that context. 
The degrees of freedom that we follow consist then of 
the Higgs fields: 
 $\Phi_1 \Phi_2 \cdot \cdot \cdot \Phi_5, \Phi_6$. It will be sufficient 
to consider them as functions of time only
which corresponds to matrix QM. This represents a sector of the full theory that 
we study presently.
 Two of the Higgs matrices
$(\Phi_5,\Phi_6 )\,\,$ are chosen to play a special 
role in the BMN scheme, defining the light cone momenta $Z = \Phi_5 + i \Phi_6 
$.  
We denote the other complex Higgs as $\, Y_i \, i=1,2$.  
In the spirit of collective field theory, the structure of 
interactions will come from considering the free $d=1$ Hamiltonian

\eqn\MatH
{H = \Tr \left( - {\partial\over\partial\bar{Z}} \, {\partial\over 
 \partial Z} + \mu^2 \bar{Z} Z \right) +\Tr 
 \left( - {\partial\over\partial \bar{Y_i}} \, {\partial 
 \over\partial Y_i} + \mu^2 \bar{Y_i} Y_i\right),}

\noindent 
since we are interested only in string coupling constant ($={1\over N}$) 
interactions. 

\subsec{Supergravity Amplitudes}

Our basic loop variables are

\eqn\Loops
{\eqalign{O^J&=\Tr (Z^J),\qquad \bar{O}^{J}=\Tr (\bar{Z}^J),\cr
\Pi^J&={\partial\over\partial O^J},\qquad
\bar{\Pi}^J={\partial\over\partial\bar{O}^J}.}}

\noindent
The two point function of our loops is

$$\langle O^J\bar{O}^J\rangle =JN^J \Big({1\over 2\mu}\Big)^J.$$

\noindent
We rescale our collective field variables 

$$\Pi^{\prime J}=\sqrt{J\Big({N\over 2\mu}\Big)^J}\Pi^J,\qquad
O^{J}=\sqrt{J\Big({N\over 2\mu}\Big)^J}O^{\prime J},$$

\noindent
so that they have a normalized two point function. Our strategy
is to consider the contribution to the collective field theory
Hamiltonian arising from the ($\Pi\Omega\Pi$) loop joining term. 
Experience from the toy model suggests this is enough to reproduce 
the cubic vertex, up to the prefactor. The loop joining contribution is

\eqn\Joins
{\eqalign{T&=\sum_{J_1, J_2}J_1 J_2\Tr (Z^{J_1-1}\bar{Z}^{J_2-1})
\Pi^{J_1}\bar{\Pi}^{J_2}\cr
&\equiv \sum_{J_1, J_2}\Omega (J_1,J_2)\Pi^{J_1}\bar{\Pi}^{J_2}.}}

\noindent
It is not surprising that the joining process has produced a loop which
is not among the variables we consider. To proceed further, we need to
express $\Omega (J_1,J_2)$ in terms of the loops we consider. 

We could imagine expanding $\Omega (J_1,J_2),$ as an operator, about its large 
$N$ value. 
The leading term in this expansion would give the contribution to the 
quadratic Hamiltonian. It is a simple task to evaluate the expectation value

\eqn\VEV
{\langle \Tr (Z^{J_1-1}\bar{Z}^{J_2-1})\rangle =\delta_{J_1 J_2}
N^{J_1}\Big({1\over 2\mu}\Big)^{J_1-1}\Big(1+O(N^{-2})\Big).}

\noindent
Using the leading term and expressing $T$ in terms of the rescaled variables, we 
obtain

$$T=\sum_J 2J\mu \Pi^{\prime J}\bar{\Pi}^{\prime J}.$$

\noindent
This is consistent with the exact spectrum of chiral loops, if we assume no 
mixing at 
the quadratic level. 

We have seen that collective field theory generates a new set of loops.
These contain both $Z$s and $\bar{Z}$s and their expectation (or classical)
value was already needed for quadratic fluctuations. More completely we  
factorize these loops into a product containing our basic (holomorphic) 
loops and the ones with nonzero classical expectation values. The 
factorization is achieved through the loop splitting formula \littleo. 
Indeed one can write

$${\partial\over\partial Z_{ij}}
(Z^{J_1-1}\bar{Z}^{J_2-2})_{ij} = {\partial\over\partial Z_{ij}}
\Big(
(Z^{J_1-1}\bar{Z}^{J_2-2})_{lk} {\partial\bar{Z}_{kl}\over\partial \bar{Z}_{ji}}
\Big) = 
\sum_{J_3=0}^{J_1-2}
Tr (Z^{J_1-2-J_3}\bar{Z}^{J_2-2})\Tr(Z^{J_3})
$$

\noindent
and the splitting is obtained as the derivative with respect to $Z_{ij}$ acts 
through the loop. This splitting rule is also the main ingredient in the 
Schwinger-Dyson equations for matrix models. It  
is known \rJStwo\ that these equations follow from the equations
of motion of this effective collective field theory action. 
This leads us to the factorization equation

\eqn\Ident
{\Tr (Z^{J_1 -1}(t)\bar{Z}^{J_2 -1}(t))\Rightarrow {A_{J_1J_2}\over 
2\mu}\sum_{J_3=0}^{J_1-2}
\Tr (Z^{J_1-2-J_3}(t)\bar{Z}^{J_2-2}(t))O^{J_3}(t),}

\noindent
for the composite loops. We also have

\eqn\IdentTw
{\Tr (Z^{J_2 -1}(t)\bar{Z}^{J_1 -1}(t))\Rightarrow {B_{J_1 J_2}\over 
2\mu}\sum_{J_3=0}^{J_1-2}
\Tr (\bar{Z}^{J_1-2-J_3}(t)Z^{J_2-2}(t))\bar{O}^{J_3}(t),}

\noindent
which can be derived in a similar way. The constant factors $A_{J_1J_2}$,
$B_{J_1 J_2}$ in above formulas represent 
normalization and can be fixed
by looking at the expectation values of both sides of this relation. The method
of taking expectations of both sides can itself be used to fully specify the 
index structure
of the form factor appearing in the above formula(see \rDDJR\ ).
Note that the large $N$ expectation values of 
$O^J$ is zero and that of $\Tr (Z^{J_1-2-J_3}(t)\bar{Z}^{J_2-2}(t))$ is given in
\VEV. Expanding each operator in \Ident\ and \IdentTw\  
about their leading large $N$ value, then to linear order in the fluctuations
we obtain 

\eqn\Ansatz
{\eqalign{\Tr (Z^{J_1 -1}\bar{Z}^{J_2 -1})&= 
N^{J_1}\Big({1\over 2\mu}\Big)^{J_1-1}\delta_{J_1 J_2}
+{J_2}\sum_{J_3=1}^{J_1-2}\delta_{J_2+J_3,J_1}N^{J_2-1}
\Big({1\over 2\mu}\Big)^{J_2-1}O_{J_3}\cr
&+{J_1}\sum_{J_3=1}^{J_2-2}\delta_{J_1+J_3,J_2}N^{J_1-1}
\Big({1\over 2\mu}\Big)^{J_1-1}\bar{O}_{J_3} .}}

\noindent
We will take this as an ansatz for the structure of $\Omega (J_1,J_2)$ to linear 
order in 
the fluctuations. Using this ansatz, we find that the loop joining contribution 
to the 
cubic vertex is

\eqn\CuV
{2\mu {\sqrt{J_1 J_2 J_3}\over N}
\big({J_1}\bar{O}^{\prime J_3} \Pi^{\prime J_1}\bar{\Pi}^{\prime 
J_2}\delta_{J_3+J_1,J_2}+{J_2} 
O^{J_3} \Pi^{\prime J_1}\bar{\Pi}^{\prime J_2}\delta_{J_3+J_2,J_1}\big).}

\noindent
As we have discussed in the example of sect.3 the transition to light cone 
fields is achieved through replacement of canonical (conjugate) fields by 
creation (annihilation) fields.
In addition we expect analogous contribution from the collective potential so 
that the final form of the cubic interaction becomes:

$$\eqalign{ H_3 = &{\sqrt{J_1 J_2 J_3}\over N}\delta_{J_1,J_2+J_3}
\Big({J_2\over J_1}\Big)^{n_2\over 2}
\Big({J_3\over J_1}\Big)^{n_3\over 2}\left(J_1 - J_2 - J_3 \right) \, 
A_{J_{1}}^{\dagger} \, A_{J_{2}}\, A_{J_{3}}\cr
&+{\sqrt{J_1 J_2 J_3}\over N}\delta_{J_1,J_2+J_3}
\Big({J_2\over J_1}\Big)^{n_2\over 2}
\Big({J_3\over J_1}\Big)^{n_3\over 2}\left(J_1 - J_2 - J_3 \right) \, 
\, A_{J_{2}}^\dagger\, A_{J_{3}}^\dagger A_{J_{1}}.}$$

We would now like to consider loops with one type of impurity added. We identify $\phi$
with $Y_1$ in \MatH\ and consider the following loop variables

\eqn\OneImp
{\eqalign{O^{J}_{n}&=\sum \Tr (\phi^n Z^J),\qquad 
\bar{O}^{J}_{n}=\sum \Tr (\bar{\phi}^n \bar{Z}^J),\cr
\Pi^{J}_{n}&={\partial\over\partial O^{J}_{n}},\qquad 
\bar{\Pi}^{J}_{n}={\partial\over\partial \bar{O}^{J}_{n}}.}}

\noindent
The sums in the definition of the loops run over all possible orderings of $n$
$\phi$s and $J$ $Z$s, i.e. there are ${(n+J)!\over n!J!}$ terms in the above 
sums.
We will work in the BMN limit\rBMN, so that $J>>n$. In this limit, the two point
functions of our loops are

\eqn\loopstp
{\eqalign{\langle O^{J_1}_{n}\bar{O}^{J_2}_{m}\rangle 
&=\delta_{J_1 J_2}\delta_{mn}\Big({1\over 2\mu}\Big)^{J_1+n}N^{J_1+n}
(J_1+n){(J_1+n)!\over J_1! n!}\cr
&=\delta_{J_1 J_2}\delta_{mn}N^{J_1+n}{J_1^{n+1}\over n!}\Big({1\over 
2\mu}\Big)^{J_1+n}.}}

\noindent
We will again work in terms of the normalized loop variables

$$ \Pi_{n}^{\prime J}=\sqrt{ \Big({N\over 2\mu}\Big)^{J+n} {J^{n+1}\over n!}}
\Pi_{n}^J,\qquad  
O_{n}^J=\sqrt{ \Big({N\over 2\mu}\Big)^{J+n} {J^{n+1}\over n!}}O_{n}^{\prime 
J}.$$

\noindent
The loop joining contribution to the Hamiltonian is given by

\eqn\OImpH
{H=\Tr\Big(
{\partial O^{J_1}_{n_1}\over \partial Z}
{\partial \bar{O}^{J_2}_{n_2}\over \partial \bar{Z}}+
{\partial O^{J_1}_{n_1}\over \partial \phi}
{\partial \bar{O}^{J_2}_{n_2}\over \partial \bar{\phi}}\Big)
\Pi^{J_1}_{n_1}\bar{\Pi}^{J_2}_{n_2}.}

\noindent
For the manipulations which follow, it is convenient to introduce the matrix 
$P(J,n)$, 
defined to be the matrix obtained by summing the ${(n+J)!\over J!n!}$ terms 
corresponding
to all possible orderings of $J$ $Z$s and $n$ $\phi$s. In terms of $P(J,n)$ we 
have

\eqn\DerivIdent
{{\partial O^J_{n}\over \partial Z_{ij}}=(J+n)P(J-1,n)_{ji}\qquad 
{\partial O^J_{n}\over \partial \phi_{ij}}=(J+n)P(J,n-1)_{ji}.}

\noindent
We can express \OImpH\ in terms of the $P(J,n)$ matrices as

\eqn\SOImpH
{\eqalign{H&=(J_1+n_1)(J_2+n_2)\Tr\Big(P(J_1-1,n_1)\bar{P}(J_2-1,n_2)\cr
&\quad +P(J_1,n_1-1)\bar{P}(J_2,n_2-1)\Big)
\Pi^{J_1}_{n_1}\bar{\Pi}^{J_2}_{n_2}\cr
&\equiv\Omega (J_1,n_1,J_2,n_2)\Pi^{J_1}_{n_1}\bar{\Pi}^{J_2}_{n_2}.}}

\noindent
In the large $N$ limit we have

\eqn\PVev
{\langle \Tr\big( P(J_1,n_1)\bar{P}(J_2,n_2)\big)\rangle =
\delta_{J_1 J_2}\delta_{n_1 n_2}N^{J_1+n_1+1}{1\over (2\mu)^{J_1+n_1}}
{(n_1+J_1)!\over n_1! J_1!}.}

\noindent
Proceeding as we did for the loops with no impurities, we obtain the following
contribution to the quadratic Hamiltonian

\eqn\SpectOImp
{T=\sum_{J,n}2(J+n)\mu\Pi^{\prime J_1}_{n_1}\bar{\Pi}^{\prime J_2}_{n_2}.}

\noindent
This again matches the exact spectrum for a loop with one type of impurity provided we 
again
assume that there is no mixing between these modes at quadratic level. To obtain 
the
cubic vertex, we need the analog of \Ident,\IdentTw\ to motivate an ansatz for 
$\Omega (J_1,n_1,J_2,n_2)$. To obtain the relevant
Schwinger-Dyson equation, we will need to use the identities 

\eqn\IdentFrP
{\eqalign{P(J,m) &=P(J-1,m)Z+P(J,m-1)\phi\cr
{\partial P(J,m)_{ij}\over\partial Z_{kl}}&=\sum_{r=0}^m\sum_{S=0}^{J-1}
P(S,r)_{ik}P(J-S-1,m-r)_{lj}\cr
{\partial P(J,m)_{ij}\over\partial \phi_{kl}}&=\sum_{r=0}^{m-1}\sum_{S=0}^{J}
P(S,r)_{ik}P(J-S,m-r-1)_{lj}.}}

\noindent
Using these identities, we are lead to the following splitting formula for the 
composite operators

\eqn\oneimp
{\eqalign{ &\Tr (P(J_1-1,n_1)\big[\bar{P}(J_2-2,n_2)\bar{Z}+\bar{P}(J_2-1,n_2-1)
\bar{\phi}\big])\cr
&+\Tr (P(J_1,n_1-1)\big[\bar{P}(J_2-1,n_2-1)\bar{Z}
+\bar{P}(J_2,n_2-2)\bar{\phi}\big])\cr
&\Rightarrow {1\over 2\mu}\sum_{r=0}^{n_1}\sum_{S=0}^{J_1-2}\Tr (P(S,r))
\Tr (P(J_1-S-2,n_1-r)\bar{P}(J_2-2,n_2))\cr 
&+{1\over \mu}\sum_{r=0}^{n_1-1}\sum_{S=0}^{J_1-1}\Tr(P(S,r))
\Tr (P(J_1-S-1,n_1-r-1)\bar{P}(J_2-1,n_2-1))\cr
&+{1\over 2\mu}\sum_{r=0}^{n_1-2}\sum_{S=0}^{J_1}\Tr(P(S,r))
\Tr (P(J_1-S,n_1-r-2)\bar{P}(J_2,n_2-2)).}}

\noindent
Noting that

$$ O^J_n=\Tr (P(J,n),$$

\noindent
the identity \oneimp\ suggests the following ansatz for $\Omega 
(J_1,n_1,J_2,n_2)$

$$\eqalign{\Omega (J_1,n_1,J_2,n_2)&=
\delta_{J_1,J_2}\delta_{n_1,n_2}(J_1+n_1)(J_2+n_2){(J_1+n_1)!\over J_1!n_1!}
{N^{J_1+n_1+1}\over (2\mu )^{J_1+n_1}}\cr
&+\delta_{J_1,J_2-J_3}\delta_{n_1,n_2-n_3}(J_1+n_1)^2(J_2+n_2){(J_1+n_1)!\over 
J_1!n_1!}
\Big({N\over 2\mu}\Big)^{J_2+n_2}
\bar{O}^{J_3}_{n_3}\cr
&+\delta_{J_1-J_3,J_2}\delta_{n_1-n_3,n_2}(J_1+n_1)(J_2+n_2)^2{(J_2+n_2)!\over 
J_2!n_2!}
\Big({N\over 2\mu}\Big)^{J_1+n_1}
O^{J_3}_{n_3}.}$$

\noindent
Using this ansatz, the loop joining contribution to the cubic vertex becomes

\eqn\CubicOnImp
{\eqalign{&{\sqrt{J_1 J_2 J_3}\over N}2\mu\sqrt{n_1!\over n_2! n_3!}
\Big({J_2\over J_1}\Big)^{n_2\over 2}
\Big({J_3\over J_1}\Big)^{n_3\over 2} (J_2 + n_2) 
\delta_{J_1,J_2+J_3}\delta_{n_1,n_2+n_3}
O_{n_3}^{\prime J_3}\Pi_{n_1}^{\prime J_1}\bar{\Pi}_{n_2}^{\prime J_2}\cr
&\quad +{\sqrt{J_1 J_2 J_3}\over N}2\mu\sqrt{n_2!\over n_1! n_3!}
\Big({J_1\over J_2}\Big)^{n_1\over 2}
\Big({J_3\over J_2}\Big)^{n_3\over 2} (J_1 + n_1 ) 
\delta_{J_2,J_1+J_3}\delta_{n_2,n_1+n_3}
\bar{O}_{n_3}^{\prime J_3}\Pi_{n_1}^{\prime J_1}\bar{\Pi}_{n_2}^{\prime J_2}.}}

As discussed previously, the transition to light cone fields is 
achieved through replacement of canonical (conjugate) fields by 
creation (anihilation) fields. Again adding the
analogous contribution from the collective potential 
the final form of the cubic interaction becomes

$$\eqalign{H_3&= {\sqrt{J_1 J_2 J_3}\over N}\sqrt{n_1!\over n_2! n_3!}\delta_{J_1,J_2+J_3}
\delta_{n_1,n_2+n_3}\Big({J_2\over J_1}\Big)^{n_2\over 2}
\Big({J_3\over J_1}\Big)^{n_3\over 2} \cr 
&\times\left( (J_1 + n_1 ) - (J_2 + n_2 ) - (J_3 + 
n_3 ) \right)
(A_{J_{1},n_{1}}^{\dagger} \, A_{J_{2} , n_{2} }\, A_{J_{3} , n_{3} }+
A_{J_{2},n_{2}}^{\dagger}\, A_{J_{3},n_{3}}^{\dagger}A_{J_{1},n_{1}}).}$$

\noindent
in agreement with the cubic interaction matrix elements for bosonic 
supergravity modes, as given for example  in \rLMP.

The above argument can be extended to loops that include more than just one 
impurity.
In order to illustrate this point, we will give the generalization to two 
impurities.
Identifying $\phi$ and $\psi$ with $Y_1$ and $Y_2$ of \MatH\ respectively, we 
consider  
the following loop variables 

\eqn\TwoImp
{O_{n,m}^J=\sum\Tr\big(\phi^n\psi^m Z^J\big) .}

\noindent
The sum on right hand side again runs over all possible orderings of the matrix 
fields, i.e.
there are ${(n+m+J)!\over n! m! J!}$ terms in the sum. The two point 
functions of our loops are

\eqn\TwoPntTwoImp
{\eqalign{\langle O^{J_1}_{n_1,m_1}&\bar{O}^{J_2}_{n_2,m_2}\rangle =
\delta_{J_1 J_2}\delta_{m_1 m_2}\delta_{n_1 n_2}
\Big({N\over 2\mu}\Big)^{J_1+n_1+m_1}\cr
&(J_1+n_1+m_1)
{(J_1+n_1+m_1)!\over J_1!n_1!m_1!}.}}

\noindent
The operators with unit two point function are obtained by rescaling

$$ \Pi^{\prime J_1}_{n_1,m_1}=\sqrt{
\Big({N\over 2\mu}\Big)^{J_1+n_1+m_1}(J_1+n_1+m_1)
{(J_1+n_1+m_1)!\over J_1!n_1!m_1!}}
\Pi^{J_1}_{n_1,m_1},$$

$$ O^{J_1}_{n_1,m_1}=\sqrt{
\Big({N\over 2\mu}\Big)^{J_1+n_1+m_1}(J_1+n_1+m_1)
{(J_1+n_1+m_1)!\over J_1!n_1!m_1!}}
O^{\prime J_1}_{n_1,m_1}.$$

\noindent
It is again convenient to introduce the matrix $M(J,n,m)$ which is a sum over 
the
${(J+n+m)!\over J!m!n!}$ terms obtained by constructing all possible 
arrangements
of $J$ $Z$ fields, $m$ $\psi$ fields and $n$ $\phi$ fields. In terms of 
$M(J,n,m)$
we have

$${\partial O_{n,m}^J\over\partial Z_{ij}}=(J+m+n)M(J-1,n,m)_{ji},$$

$${\partial O_{n,m}^J\over\partial \phi_{ij}}=(J+m+n)M(J,n-1,m)_{ji},$$

$${\partial O_{n,m}^J\over\partial \psi_{ij}}=(J+m+n)M(J,n,m-1)_{ji},$$

\noindent
As before, we consider only the loop joining contribution to the collective 
field 
theory Hamiltonian

$$\eqalign{H&=\Omega (J_1,n_1,m_1,J_2,n_2,m_2)
\Pi^{J_1}_{n_1,m_1}\bar{\Pi}^{J_2}_{n_2,m_2}\cr
=\Big(&{\partial O_{n_1,m_1}^{J_1}\over\partial Z_{ij}}
{\partial \bar{O}_{n_2,m_2}^{J_2}\over\partial \bar{Z}_{ji}}+
{\partial O_{n_1,m_1}^{J_1}\over\partial \phi_{ij}}
{\partial \bar{O}_{n_2,m_2}^{J_2}\over\partial \bar{\phi}_{ji}}+
{\partial O_{n_1,m_1}^{J_1}\over\partial \psi_{ij}}
{\partial \bar{O}_{n_2,m_2}^{J_2}\over\partial \bar{\psi}_{ji}}\Big)
\Pi^{J_1}_{n_1,m_1}\bar{\Pi}^{J_2}_{n_2,m_2}.}$$

\noindent
Using the two point function

$$\eqalign{\langle\Tr\big( M&(J_1,n_1,m_1)\bar{M}(J_2,n_2,m_2)\big)\rangle =
\delta_{J_1 J_2}\delta_{m_1 m_2}\delta_{n_1 n_2}\cr
&\times {(J_1+n_1+m_1)!\over J_1!n_1!m_1!}
{N^{J_1+n_1+m_1+1}\over (2\mu)^{J_1+n_1+m_1}}}$$

\noindent
we find the following contribution to the quadratic Hamiltonian

$$ H=(J+n+m) \Pi^{\prime J}_{n,m}\bar{\Pi}^{\prime J}_{n,m}.$$

\noindent
To obtain the ansatz for $\Omega (J_1,n_1,m_1,J_2,n_2,m_2)$ needed to obtain
the cubic vertex, we use the identities

$$ M(J,n,m)=M(J-1,n,m)Z+M(J,n-1,m)\phi +M(J,n,m-1)\psi ,$$

$${\partial M(J,n,m)_{ij}\over \partial Z_{kl}}=
\sum_{r=0}^{J-1}\sum_{s=0}^n\sum_{t=0}^m
M(r,s,t)_{ik}M(J-1-r,n-s,m-t)_{lj},$$

$${\partial M(J,n,m)_{ij}\over \partial \phi_{kl}}=
\sum_{r=0}^{J}\sum_{s=0}^{n-1}\sum_{t=0}^m
M(r,s,t)_{ik}M(J-r,n-s-1,m-t)_{lj},$$

$${\partial M(J,n,m)_{ij}\over \partial \psi_{kl}}=
\sum_{r=0}^{J}\sum_{s=0}^n\sum_{t=0}^{m-1}
M(r,s,t)_{ik}M(J-r,n-s,m-t-1)_{lj},$$

\noindent
We obtain the following identity between $d=1$ operators

\vfill\eject

$$\eqalign{&\Tr\Big(
M(J_1-1,n_1,m_1)\bar{M}(J_2-1,n_2,m_2)+
M(J_1,n_1-1,m_1)\bar{M}(J_2,n_2-1,m_2)+\cr
&+M(J_1,n_1,m_1-1)\bar{M}(J_2,n_2,m_2-1)\Big)\cr
&\Rightarrow
{1\over 2\mu}\sum_{J=0}^{J_1-2}\sum_{n=0}^{n_1}\sum_{m=0}^{m_1}
\Tr (M(J,n,m))
\Tr\Big(M(J_1-2-J,n_1-n,m_1-m)\bar{M}(J_2-2,n_2,m_2)\Big)\cr
&+{1\over \mu}\sum_{J=0}^{J_1-1}\sum_{n=0}^{n_1-1}\sum_{m=0}^{m_1}
\Tr (M(J,n,m))
\Tr\Big(M(J_1-1-J,n_1-n-1,m_1-m)\cr
&\qquad\times\bar{M}(J_2-1,n_2-1,m_2)\Big)\cr
&+{1\over \mu}\sum_{J=0}^{J_1-1}\sum_{n=0}^{n_1}\sum_{m=0}^{m_1-1}
\Tr (M(J,n,m))
\Tr\Big(M(J_1-1-J,n_1-n,m_1-m-1)\cr
&\qquad\times\bar{M}(J_2-1,n_2,m_2-1)\Big)\cr
&+{1\over 2\mu}\sum_{J=0}^{J_1}\sum_{n=0}^{n_1-2}\sum_{m=0}^{m_1}
\Tr (M(J,n,m))
\Tr\Big(M(J_1-J,n_1-n-2,m_1-m)\bar{M}(J_2,n_2-2,m_2)\Big)\cr
&+{1\over \mu}\sum_{J=0}^{J_1}\sum_{n=0}^{n_1-1}\sum_{m=0}^{m_1-1}
\Tr (M(J,n,m))
\Tr\Big(M(J_1-J,n_1-n-1,m_1-m-1)\cr
&\qquad\times\bar{M}(J_2,n_2-1,m_2-1)\Big)\cr
&+{1\over 2\mu}\sum_{J=0}^{J_1}\sum_{n=0}^{n_1}\sum_{m=0}^{m_1-2}
\Tr (M(J,n,m))\Tr\Big(M(J_1-J,n_1-n,m_1-m-2)\bar{M}(J_2,n_2,m_2-2)\Big).}$$

\noindent
Using this equation to motivate an ansatz in our usual fashion, we land up with 
the following loop joining contribution to the
cubic vertex

\eqn\TwoImp
{\eqalign{&2\mu\delta_{J_1,J_2+J_3}\delta_{n_1,n_2+n_3}\delta_{m_1,m_2+m_3}
{\sqrt{J_1 J_2 J_3}\over N}\sqrt{n_1!\over n_2! n_3!}
\sqrt{m_1!\over m_2! m_3!}\cr
&\quad\times \left( J_2 + n_2 + m_2 \right) \Big({J_2\over J_1}\Big)^{n_2+m_2\over 2}
\Big({J_3\over J_1}\Big)^{n_3+m_3\over 2}
\Pi^{\prime J_1}_{n_1,m_1}
\bar{\Pi}^{\prime J_2}_{n_2,m_2}
O^{\prime J_3}_{n_3,m_3}.}}

\noindent
This again implies a light cone coupling of the form

$$ \left( \Delta J + \Delta n + \Delta m\right)
\sqrt{n_1!\over n_2! n_3!}\sqrt{m_1!\over m_2! m_3!}
\delta_{J_1,J_2+J_3}\delta_{n_1,n_2+n_3}\delta_{m_1,m_2+m_3}
\Big({J_2\over J_1}\Big)^{n_2+m_2\over 2}
\Big({J_3\over J_1}\Big)^{n_3+m_3\over 2}$$

\noindent 
with $\, \Delta J = J_1 + J_2 - J_3 \,$ and similarly for $\, n\,$ and 
$\, m \,$. This is in agreement with the cubic interaction matrix elements for bosonic 
supergravity modes.

\newsec{Lattice string}

In this section we show that in position (lattice) space the collective field
theory produces the  ultralocal vertex of string field theory.

Our general collective field variables are now given by

\eqn\LattStr
{\Phi_J(\{ l\})=\Tr\Big( T_l\prod_{i=1}^n\phi (l_i) Z^J\Big),\qquad
\phi (l_i )=Z^{l_i}\phi Z^{-l_i},}

\noindent
with $T_l$ the $l$ ordering operator - it orders the $\phi (l)$ factors so that 
$l_i$ increases 
from left to right. As an example,
the correspondence between collective fields and states in the ultralocal 
string field theory Hilbert space is given by

\eqn\Corr
{\Phi_J(\{l_i, l_i, l_i, l_j, l_j\})\leftrightarrow
{(a^\dagger(i))^3\over \sqrt{3!}}{(a^\dagger(j))^2\over \sqrt{2!}}|0,J\rangle .}

\noindent
We also have

\eqn\ConjFld
{\bar{\Phi}_J (\{ l\})=\Tr\Big( \bar{Z}^J\tilde{T}_l\prod_{i=1}^n \bar{\phi} 
(l_i) \Big),\qquad
\bar{\phi} (l_i )=\bar{Z}^{-l_i}\bar{\phi} \bar{Z}^{l_i}.}

\noindent 
$\tilde{T}_l$ is a second $l$ ordering operator - it orders the $\bar{\phi}(l)$ 
factors so that 
$l_i$ {\it decreases} from left to right. The loops are orthogonal 
at large $N$, i.e.

$$\langle \Phi_J (\{ k\})\bar{\Phi}_{\bar{J}} (\{ l\})=\delta_{J\bar{J}}
\prod_{i=1}^n\delta_{k_i l_i}{1\over (2\mu )^{J+n}}.$$

\noindent
Introduce  

$$ P_{J}(\{ l\}))_{ij}=\sum_{a=1}^J T_l\Big(\Big[\prod_{i=1}^n \phi 
(l_i-a)Z^{J-1}\Big]_{ij}\Big)
= {\partial \Phi_J(\{ l\})\over\partial Z_{ji}}$$  

\noindent 
In the last formula we take $l_i-a$ mod $J$ so that $0\le l_i-a\le J-1$. 
Introduce  

$$ Q_{J}(\{ l\}))_{ij}=\sum_{j=1}^n T_l\Big(\prod_{i=1,i\ne j}^n
\Big[\phi (l_i-l_j)Z^{J}\Big]_{ij}\Big)= 
{\partial \Phi_J(\{ l\})\over\partial \phi_{ji}}$$  

\noindent 
In the last formula we take $l_i-l_j$ mod $J$. At leading order in $N$ we have  

$$\langle \Tr\Big[T_l\Big(\prod_{i=1}^n\phi 
(l_i)\Big)Z^J\bar{Z}^{\bar{J}}\tilde{T}_l
\Big( \prod_{j=1}^{\bar{n}}\bar{\phi}(\bar{l}_j)\Big) 
\Big]\rangle=\delta_{n\bar{n}}\delta_{J\bar{J}}N^{n+J+1}
\prod_{i=1}^n\delta_{l_i\bar{l}_i}{1\over (2\mu )^{J+n}}. $$  

\noindent 
Next, we need to study the quantity  

$$\eqalign{ {\partial\Phi_{J_1}(\{ l\})\over\partial Z_{ij}} 
&{\partial\bar{\Phi}_{J_2}(\{ \bar{l}\})\over\partial\bar{Z}_{ji}}
+ {\partial\Phi_{J_1}(\{ l\})\over\partial \phi_{ij}} 
{\partial\bar{\Phi}_{J_2}(\{ \bar{l}\})\over\partial\bar{\phi}_{ji}}\cr 
&=\Tr (P_{J_1}(\{ l\})\bar{P}_{J_2}(\{\bar{l}\})) +\Tr (Q_{J_1}(\{ 
l\})\bar{Q}_{J_2}(\{\bar{l}\}))\cr 
&=\sum_{a=1}^{J_1}\sum_{b=1}^{J_2}\Tr\Big[ T_l\Big(\prod_{i=1}^{n_1}\phi 
(l_i-a)\Big)Z^{J_1-1} 
\bar{Z}^{J_2-1}\tilde{T}_l\Big(\prod_{j=1}^{n_2}\bar{\phi}(\bar{l}_j-b)\Big)\Big
]\cr 
&+\sum_{k=1}^{n_1}\sum_{m=1}^{n_2}\Tr\Big[ T_l\Big(\prod_{i=1,i\ne k}^{n_1}\phi 
(l_i-l_k)\Big)Z^{J_1} 
\bar{Z}^{J_2}\tilde{T}_l\Big(\prod_{j=1,j\ne 
m}^{n_2}\bar{\phi}(\bar{l}_j-\bar{l}_m)\Big)\Big]}$$  

\noindent 
 It is helpful to 
introduce  

$$ P'_{J}(\{ l\}))_{ij}=\sum_{\matrix{a=1,\cr l_i-a\ne J-1}}^J 
T_l\Big(\prod_{i=1}^n
\Big[\phi (l_i-a)Z^{J-2}\Big]_{ij}\Big)$$  $$\tilde{P}_{J}(\{ 
l\}))_{ij}=\sum_{j=1}^n T_l
\Big(\prod_{\matrix{i=1,\cr i\ne j}}^n\Big[\phi 
(l_i-l_j-1)Z^{J-1}\Big]_{ij}\Big)$$  

\noindent 
To get $P_J'$ from $P_J$, keep all terms that end with a $Z$ and strip off the 
last $Z$. To get 
$\tilde{P}_J$ from $P_J$, keep all terms ending with a $\phi$ and strip off the 
last $\phi$.  

$$ P'_{J}(\{ l\})Z+\tilde{P}_{J}(\{ l\})\phi =P_{J}(\{ l\}) $$  

\noindent 
It is straight forward to verify that  

$$\eqalign{{\partial P_{J_1}(\{ l\})_{kl}\over\partial 
Z_{ij}}&=\sum_{a=1}^{J_1}\sum_{J_3=0}^{J_1-1} 
\Big( T_l\Big[\prod_{\matrix{i=1\cr l_i-a\le J_3\cr l_i-a\ne J_1-1}}^n \phi (l_i 
-a)Z^{J_3}\Big]_{ki}\Big)\cr 
&\times \Big(T_l\Big[ \prod_{\matrix{i=1\cr l_i-a\ge J_3 +1\cr l_i-a\ne 
J_1-1}}^n\phi (l_i-a-J_3-1) 
Z^{J_1-J_3-2}\Big]_{jl}\Big)}$$  

\noindent 
The analysis of the supergravity modes given in the previous section can be used 
to argue that
we can drop 
$\Tr ({\partial\over\partial\phi}{\partial\over\partial\phi})$ 
and keep just the $\Tr ({\partial\over\partial 
Z}{\partial\over\partial\bar{Z}})$ term. 
For this reason, concentrate on  

$$\eqalign{{\partial\Phi_{J_1}(\{ l\})\over\partial Z_{ij}} 
&{\partial\bar{\Phi}_{J_2}(\{ l\})\over\partial\bar{Z}_{ji}} \cr
&=\sum_{a=1}^{J_1}\sum_{b=1}^{J_2}\Tr\Big[ T_l\Big(\prod_{i=1}^{n_1}\phi 
(l_i-a)\Big)Z^{J_1-1} 
\bar{Z}^{J_2-1}\tilde{T}_l
\Big(\prod_{j=1}^{n_2}\bar{\phi}(\bar{l}_j-b)\Big)\Big],}$$

\noindent 
writting it as 

$$\eqalign{\Tr &\big(P_{J_1}(\{l\})\bar{P}_{J_2}(\{ \bar{l}\})\big)=
{\partial\over\partial Z_{ij}} 
\Big( \big[ P_{J_1}(\{l\})\bar{P}'_{J_2}(\{ \bar{l}\})\big]_{ij}\Big)\cr 
&+{\partial\over\partial \phi_{ij}}
\Big( \big[ P_{J_1}(\{l\})\bar{\tilde{P}}_{J_2}(\{ \bar{l}\})\big]_{ij}\Big) 
.}$$  

\noindent 
Keeping only the leading term we obtain

\vfill\eject
  
$$\eqalign{\Tr \big(P_{J_1}(\{l\})&\bar{P}_{J_2}(\{ \bar{l}\})\big) 
\Rightarrow {1\over 2\mu}\sum_{a=1}^{J_1}\sum_{J_3=0}^{J_1-1}
\Tr\Big( T_l\Big[\prod_{\matrix{i=1\cr l_i-a\le J_3\cr l_i -a\ne J_1-1}}^{n_3}
\phi (l_i-a)Z^{J_3}\Big]\Big)\cr 
&\times \Tr\Big( T_l\Big[
\prod_{\matrix{i=1\cr l_i-a\ge J_3 +1\cr l_i-a\ne J_1-1}}^{n_1-n_3}
\phi (l_i -a-J_3-1)\big]Z^{J_1-J_3-2} \bar{P}'_{J_2}(\{\bar{l}\})\Big)\cr 
&={1\over 2\mu}\sum_{a=1}^{J_1}\sum_{J_3=0}^{J_1-1}
\Tr\Big( T_l\big[\prod_{i=1}^{n_3}\phi(p_i)Z^{J_3}\big]\Big)\times\cr
&\prod_{\matrix{i=1\cr l_i-a\le J_3\cr l_i -a\ne J_1-1}}^{n_3}\delta_{p_i,l_i-a}
\prod_{\matrix{i=1\cr l_i-a\ge J_3 +1\cr l_i-a\ne J_1-1}}^{n_1-n_3}
\delta_{q_i,l_i-a-J_3-1}\cr
&\times\Tr\Big( T_l\big[\prod_{i=1}^{n_1-n_3}\phi(q_i)\big]
Z^{J_1-J_3-2} \bar{Z}^{J_2-2}\tilde{T}_l\big[\prod_{i=1}^{n_2}
\bar{\phi}(\bar{l}_i)\big]\Big)\cr 
&={1\over 
2\mu}\sum_{a=1}^{J_1}\sum_{J_3=0}^{J_1-1}\delta_{\{p\},\{l\}}\delta_{\{q\},\{l\}
}\cr
&\Phi_{J_3}(\{ p\})\Tr\Big( 
T_l\big[\prod_{i=1}^{n_3-n_2}\phi(q_i)\big]Z^{J_1-J_3-2} 
\bar{Z}^{J_2-2}\tilde{T}_l\big[
\prod_{i=1}^{n_2}\bar{\phi}(\bar{l}_i)\big]\Big)}$$  

\noindent 
After rescaling and normalization, our cubic 
interaction takes the form

$$ {2\mu\over N} (J_1 + n_1) \delta_{J_1-J_3,J_2}\delta_{n_1-n_3,n_2}
\prod_{i=1}^{n_2}\delta_{q_i,\bar{l}_i}
\prod_{j=1}^{n_3}\delta_{p_{j}q_{j+n_2}} 
\Phi_{J_3}(\{ p\})\Pi_{J_1}(\{q\})\bar{\Pi }_{J_2}(\{\bar{l}\}) $$  

\noindent
 Again in the light cone limit and with the matching contribution coming from 
the collective potential, we are lead to the following 3-string interaction:

$$
\eqalign{ & H_3^{col}=\cr & \sum_{J_1,J_2,J_3}
\sum_{\{l^{(1)}\},\{l^{(2)}\},\{l^{(3)}\}} \left(\Delta J+\Delta n \right) 
\langle\psi_{1}| \, \langle\psi_{2}| \, \langle\psi_{3}| 
\,
|V_{3}^{0}\rangle \Phi_{J_3}(\{l^{(3)}\}) {\partial\over \partial 
\Phi_{J_1}(\{l^{(1)}\})} {\partial\over \partial 
\Phi_{J_2}(\{l^{(2)}\})} \cr
& +h.c.}
$$
Here the prefactor again contains the energies in the form $(\, E_3^0 - E_1^0 - 
E_2^0 )\,$ and $\, E^0 = J + n\,. $ 

\noindent 
This interaction vertex of collective field theory matches up with the 
string field theory vertex. It is represented by

$$ |V\rangle =e^{\sum_{l=0}^{\alpha_1-1}a_1^\dagger (l) a_3^\dagger (l) 
+\sum_{l=\alpha_1}^{\alpha_3 -1}a_2^\dagger (l-\alpha_1)a_3^\dagger 
(l)}|0\rangle .$$

\noindent
Using the notation (we order the $l_i$'s so that $l_i\ge l_j$ if $i>j$)

$$ |\alpha_1,\{l_1, n_1 \},\{ l_2, n_2\},...\{ l_N, n_N\}\rangle\equiv 
{(a(l_1)^\dagger)^{n_1}\over n_1!}{(a(l_2)^\dagger)^{n_2}\over n_2!}...
{(a(l_N)^\dagger)^{n_N}\over n_N!}|0,\alpha_1\rangle,$$

\noindent
for a state with occupation numbers $n_i$ at sites $l_i$, we find

\eqn\LattStrAmp
{\eqalign{\langle &V|\alpha_1,\{l_1, n_1 \},...,\{ l_N, n_N\}\rangle_1
|\alpha_2,\{m_1, p_1 \},...,\{ m_M, p_M\}\rangle_2\cr
&\qquad |\alpha_3,\{q_1, r_1 \},...,\{ q_{N+M}, r_{N+M}\}\rangle_3\cr
&=\delta_{\alpha_1+\alpha_2,\alpha_3}\delta_{l_1 q_1}\delta_{n_1 r_1}
...\delta_{l_N q_N}\delta_{n_N r_N}
\delta_{m_1+\alpha_1, q_{N+1}}\delta_{p_1 r_{N+1}}\cr
&\qquad\times ...\delta_{m_M+\alpha_1, q_{N+M}}\delta_{p_M r_{N+M}}.}}
\noindent
demonstrating the agreement.

\newsec{Conclusions}

We have shown in the present work how based on the collective field 
method one generates an interacting string
field theory from the dynamics of matrices. The interactions
are seen to originate from the free (kinetic) term of a matrix theory,
they come proportional to 1/$N$. In general, gauge theory
interactions will renormalize these basic string interactions. 
The  described method offers a potential for constructing 
both types (perturbative and nonperturbative) of 
interactions. In the earlier application of the theory to noncritical
strings one has not differentiated between the two.  We 
will present this discussion in a future publication.

The method being
based on direct, time-like hamiltonian 
formalism is fundamentally unitary. That is the function of the (highly) 
nontrivial
Jacobian transformation that defines collective field theory. 
Apart from leading to a sequence of higher point  vertices it also provides 
necessary counterterms whose effect 
becomes relevant at the loop level (see \rDJR\ for examples of such 
calculations). The 
present scheme then offers a potential for direct and complete map from matrix 
theory to string field theory.

Collective string field theory (CSFT) amplitudes as a rule agree with results 
computed by other means. But it is relevant to stress that
CSFT represents a generalization and a broader framework compared to standard 
string 
field theory (SFT). While standard light-cone  SFT is set 
in a fixed background (for example the pp wave) in collective string field 
theory (CSFT) the background is  determined as a 
solution of the (collective) equation. Equivalently it 
can be specified by the evaluation of 'one-point' functions (Wilson loops) in 
Yang-Mills 
theory. The full theory exhibits a number of degrees of freedom much 
larger then that contained in standard SFT. As we have demonstrated the extra 
fields
play a relevant role in establishing the background since 
their expectation values are seen to be nonzero. 
In this very fundamental sense CSFT  represents an extension of customary SFT.

$$ $$

\noindent
{\it Acknowledgements:} 
One of us (AJ) has had many discussions on the present topic 
with Bunji Sakita. We are most
grateful for his insight and ideas that permeate this paper. 
He will be greatly missed.
The work of RdMK and JPR is supported by NRF grant number Gun 2053791.
The work of AJ is supported by DOE grant DE FGO2/19ER40688(Task A).

\listrefs
\vfill\eject
\bye